\documentclass[
preprint,
superscriptaddress,
showpacs,
amsmath,amssymb,
aps,prl,
]{revtex4-2}
\usepackage{amssymb}
\usepackage{amsmath}
\usepackage{booktabs}
\usepackage{tabularx}
\usepackage{braket}
\usepackage{float}
\usepackage{graphicx}
\usepackage{dcolumn}
\usepackage{bm}
\usepackage{subfigure}
\usepackage{color}

\usepackage{epstopdf, epsfig}
\usepackage{blindtext}
\usepackage{longtable}
\usepackage{amsmath}
\usepackage{mathtools}
\usepackage{enumitem}
\usepackage[T1]{fontenc}
\usepackage{mathptmx}

\usepackage[colorlinks,citecolor = blue, linkcolor=red,hyperindex,CJKbookmarks]{hyperref}

\begin{document}

\title{Bistability in radiatively heated melt ponds}

\author{Rui Yang}
\affiliation{Physics of Fluids Group and Max Planck Center for Complex Fluid Dynamics, and
J. M. Burgers Centre for Fluid Dynamics, University of Twente, P.O. Box 217, 7500AE Enschede,
The Netherlands}
\author{Christopher J. Howland}
\affiliation{Physics of Fluids Group and Max Planck Center for Complex Fluid Dynamics, and
J. M. Burgers Centre for Fluid Dynamics, University of Twente, P.O. Box 217, 7500AE Enschede,
The Netherlands}
\author{Hao-Ran Liu}
\affiliation{Department of Modern Mechanics, University of Science and Technology of China, Hefei 230027, China}
\author{Roberto Verzicco}
\affiliation{Physics of Fluids Group and Max Planck Center for Complex Fluid Dynamics, and
J. M. Burgers Centre for Fluid Dynamics, University of Twente, P.O. Box 217, 7500AE Enschede,
The Netherlands}
\affiliation{Dipartimento di Ingegneria Industriale, University of Rome 'Tor Vergata', Via del Politecnico 1, Roma 00133, Italy}
\affiliation{Gran Sasso Science Institute - Viale F. Crispi, 7, 67100 L'Aquila, Italy}
\author{Detlef Lohse}\email{d.lohse@utwente.nl}
\affiliation{Physics of Fluids Group and Max Planck Center for Complex Fluid Dynamics, and
J. M. Burgers Centre for Fluid Dynamics, University of Twente, P.O. Box 217, 7500AE Enschede,
The Netherlands}
\affiliation{Max Planck Institute for Dynamics and Self-Organisation, Am Fassberg 17, 37077 G{\"o}ttingen, Germany}

\date{\today}

\begin{abstract}
Melting and solidification processes, intertwined with convective flows, play a fundamental role in geophysical contexts. One of these processes is the formation of melt ponds on glaciers, ice shelves, and sea ice. It is driven by solar radiation and is of great significance for the Earth's heat balance, as it significantly lowers the albedo. Through direct numerical simulations and theoretical analysis, we unveil a bistability phenomenon in the melt pond dynamics. As solar radiation intensity and the melt pond's initial depth vary, an abrupt transition occurs: This tipping point transforms the system from a stable fully frozen state to another stable equilibrium state, characterized by a distinct melt pond depth. The physics of this transition can be understood within a heat flux balance model, which exhibits excellent agreement with our numerical results. Together with the Grossmann-Lohse theory for internally heated convection, the model correctly predicts the bulk temperature and the flow strength within the melt ponds, offering insight into the coupling of phase transitions with adjacent turbulent flows and the interplay between convective melting and radiation-driven processes.
\end{abstract}

\maketitle

Flow mediated melting and solidification are fundamental processes in geophysics, encompassing icebergs \citep{ristroph2018sculpting}, subglacial lakes, and sea ice \cite{holland2006future}. Accurately quantifying these processes is critical for improving the estimate of the global ice melt rate \cite{eisenman2009nonlinear,edwards_projected_2021,cenedese2022icebergs} and climate models, where many small-scale processes are usually disregarded or parameterized \cite{holland1999role,stroeve2007arctic}. One of these small-scale processes is the formation and evolution of melt ponds, which recently have been identified as a key factor in the energy balance of Earth's polar regions \cite{schroder2014september}, and therefore in correctly modeling climate change. The reason is that the presence of liquid water on the surface of ice significantly alters the surface albedo, leading to increased absorption of solar radiation rather than reflection \cite{holland2006future}. This enhanced absorption intensifies heating and accelerates the melting process. Consequently, the formation and growth of melt ponds globally contribute to the amplification of ice melt, including ice shelves \cite{van2023variable}, sea ice \cite{popovic2018simple}, ice sheets \cite{sneed2007evolution}, and mountain glaciers \cite{miles2018surface}. 

Previous studies considered different modeling approaches to investigate the evolution of melt ponds, such as a simple disk-filling model \cite{popovic2018simple}, a 1D melt-pond model based on a two-stream radiation mode \cite{taylor2004model}, and an Ising model for melt ponds \cite{ma2019ising}, which provides similarity with real melt ponds in the length scale and fractal dimension. However, due to their complexity and numerous strong \& simplifying assumptions, existing models of melt pond evolution provide only a limited fundamental understanding. Therefore, next to field measurement, lab experiments, and direct numerical simulations (DNS) of the melting process with radiative heating are indispensable to understanding the evolution of melt ponds from a fundamental point of view. 

Such lab experiments and DNS have often been cast in the framework of Rayleigh-Bénard (RB) convection \cite{davis1984pattern,esfahani2018basal,favier2019,yang2023morphology}, i.e., the flow in a box heated from below and cooled from above \cite{ahlers2009,Lohse2010,Chilla2012,shishkina2021rayleigh}. What must be considered here is that the water within melt ponds is typically freshwater below $4^oC$, where density increases with temperature. Indeed, this density anomaly effect on the melting process plays a significant role in the flow structure and melt dynamics \cite{wang2021growth,wang2021equilibrium,wang2021ice,yang2022abrupt}. Moreover, the influence of solar radiative heating, the primary heat source for melt ponds, which is usually considered as an exponentially decaying profile, is different from the fixed temperature or heat flux boundary condition in RB. Factors such as cloud patterns and diurnal cycles can cause solar radiation variations. Additionally, natural factors like topography or the accumulation of meltwater can lead to the formation of a water layer \cite{Scagliarini2020}. These effects are still less investigated.

\begin{figure}[ht]
 \centering
 \includegraphics[width=1.0\columnwidth]{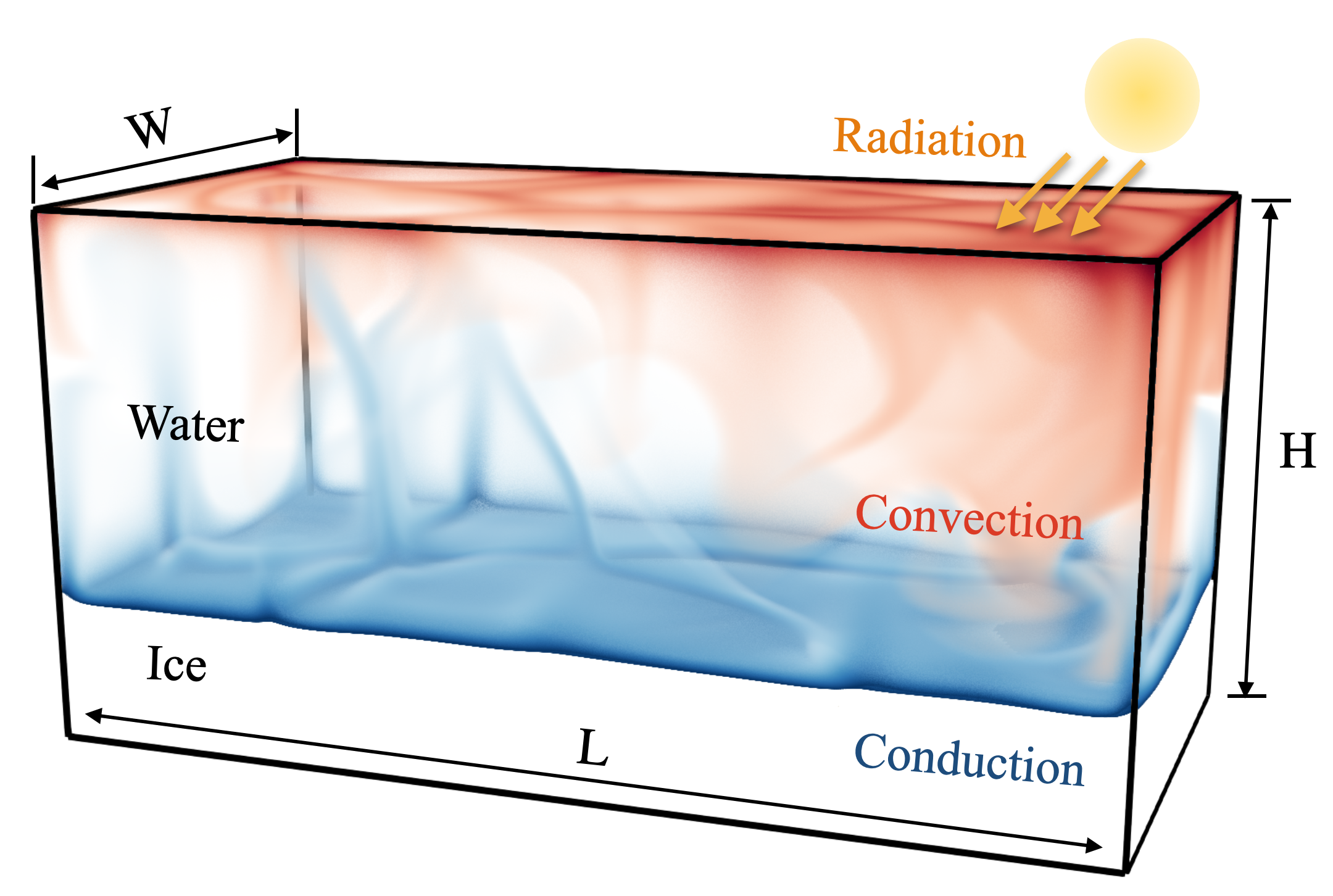}
 \caption{Snapshot of the temperature field of our simplified simulation setup at the final stable equilibrium stages with melt pond formation, with the diffusive equilibrium initial condition from eq.~(\ref{eq:Tprof}).}
 \label{fig1}
\end{figure}

The object of this Letter is a comprehensive understanding of melt pond formation. Therefore, within the RB framework, we numerically investigate the evolution of the melt pond depth under (varying) solar radiation, for which we use the Beer-Lambert model \cite{calloway1997beer}. Our simple model captures the key physical processes relevant to melt ponds: radiation, convection, and melting. Our main numerical finding is a bistability of the system, leading to a tipping-point-like behavior, where a small change in radiation strength or initial melt pond depth triggers an abrupt change in the final melt pond state. We further provide a theoretical explanation for the observed bistability and regime transition based on the heat flux balance between ice and water, exhibiting good agreement with our numerical results. Additionally, we successfully predict the dependences of the bulk temperature and the flow intensity on the solar radiation strength with the help of the Grossmann-Lohse theory for internally heated convective flows \cite{grossmann2000scaling,grossmann2001thermal,wang2021scaling}.

We perform both two-dimensional (2D) and three-dimensional (3D) numerical simulations, integrating the velocity field $\textbf{u}$ and the temperature field $\theta$ according to the Navier-Stokes equations, employing the Oberbeck-Boussinesq (OB) approximation, i.e., we consider the density $\rho$ to be linear in temperature as $\rho=\rho_0(1+\alpha(T-T_0))$, with the thermal expansion coefficient $\alpha>0$ since the maximum density of water is attained at $4^oC$ and melt ponds are typically below $2^oC$ \cite{kim2018salinity}, and assume all other material properties to be temperature independent. The melting/freezing process is simulated with the phase-field method (PF), which has been extensively used for phase boundary evolutions \cite{favier2019,hester2020improved,hester2021aspect,couston2021topography,yang2023morphology}. More details of the method and validations can be found in our previous work \cite{yang2023morphology,yang2023salinity}. We further checked the effect of grid resolution and the phase-field-related parameter on the bifurcation and found robustness, see Supplementary Material. We set the top wall as free slip and adiabatic, the bottom wall as a fixed temperature $T_c=-3^oC$ (a simplified estimated value from experiment measurement \cite{bliss2020glaciers}, in a real scenario it likely depends on heat flux through the solid ice -- the effect of $T_c$ will be discussed below), and the sidewalls as no-slip and adiabatic. Lateral heat loss could also matter, as it can modify the flow pattern and the local heat flux. Here, we neglected its effect by assuming the melt pond is wide enough and thus the sidewall effect only plays a minor role, similarly as it does for the evolution of the melt pond morphology \cite{Scagliarini2020}. Our model flow is confined in a rectangular box, with height $H$, width $W$, and length $L$, with gravity acting vertically downwards. The aspect ratio is set at $W/H=2$ for 2D and $L/H=1$ for 3D, with a dimensional depth of $H=0.2m$, representative of common melt pond depths \cite{skyllingstad2007numerical,taylor2004model}.

\begin{figure}[ht]
 \centering
 \includegraphics[width=1.0\columnwidth]{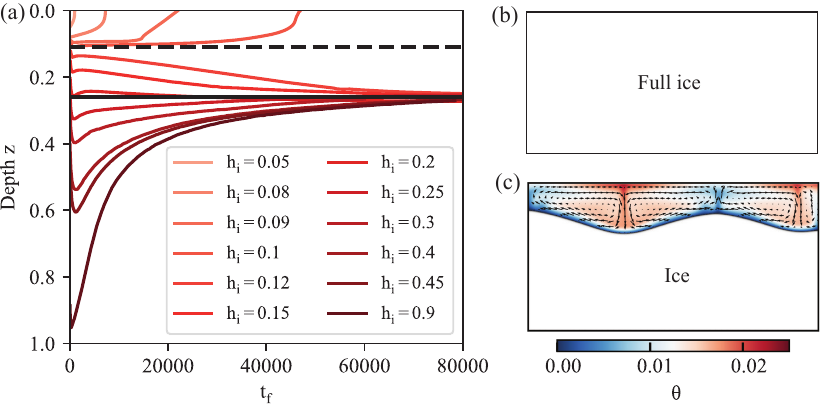}
 \caption{(a) The temporal evolution of 2D melt pond depth for various $h_i$ while maintaining a fixed $I_0=35W/m^2$. The initial temperature profile is given by eq.~(\ref{eq:Tprof}). $t_f$ is the dimensionless time in free-fall units. The solid (dashed) straight line indicates the stable (unstable) equilibrium for these parameters. (b) The equilibrium state of full ice. (c) The equilibrium state of a melt pond.}
 \label{fig2}
\end{figure}

The commonly used absorption law for solar radiation is the Beer-Lambert law \cite{calloway1997beer}. Building upon this, the parametrization \cite{skyllingstad2007numerical}
\begin{equation}
F(Z)=I_0\sum\limits_{m=1}^{4} P_m\left(1-e^{-K_m Z}\right),
\label{eq:profile}
\end{equation}
is used for the solar flux. Here $I_0$ represents the solar flux as the incident shortwave radiation at the surface and $Z$ represents the depth. The parameters $P_m$ and $K_m$ correspond to the proportion of energy in the band $m$ and the diffuse extinction coefficient, respectively (values are taken from \cite{skyllingstad2007numerical} and provided in Table I in the Supplementary Materials). In our simple model, we neglect the contribution of reflected/transmitted shortwave radiation from the bottom of the melt pond \cite{skyllingstad2007numerical}, because we want to model the physical process while maintaining a simplified setup. The primary dimensional control parameter is $I_0$, which in dimensionless form is defined as Rayleigh number $Ra$. The other dimensionless control parameters are the Prandtl number $Pr$, the Stefan number $St$, and the initial water depth $h_i$ of the water layer,
\begin{equation}
Ra=\frac{\alpha g I_0H^4}{\rho_0 c\nu\kappa^2},\ Pr=\frac{\nu}{\kappa},\ St=\frac{\cal{L}}{c\Delta_T}=\frac{\rho_0\kappa\cal{L}}{I_0H},\ h_i=\frac{H_i}{H},
\end{equation}
where $c$ is the specific heat capacity, $\nu$ the kinematic viscosity, $\kappa$ the thermal diffusivity, $\cal{L}$ the latent heat, $H_i$ the dimensional initial water depth, and $\Delta_T=(I_0H)/(\rho_0 c\kappa)$ can be seen as the characteristic dimensionless temperature difference of the system. We fix $Pr=10$ as for cold water and $St=10^{-3}$, which is smaller than the realistic value. We do so anyhow because of numerical efficiency, as smaller $St$ means faster melting/freezing to reach the equilibrium state. Since the Stefan number only appears in front of the interface velocity in the governing equations, it has no impact on the equilibrium states. Previous studies of melting convective systems have also only found a minor influence of $St$ on the flow evolution \cite{favier2019,yang2023morphology}. $St$ may transiently matter if it is so small that the melt pond quickly freezes before the convection onset occurs. We checked that the $St$ we chose has only a minor effect on the bifurcation, see Supplementary Material We investigate varying radiation strengths $6\times10^7\le Ra\le6\times10^9$, corresponding to $10^1\lesssim I_0\lesssim 10^3\ W/m^2$, and varying initial depths $0\le h_i\le1$, and examine how they affect the formation of the melt pond , where we use the horizontal averaged water depth to quantify the melt pond depth. The initial temperature profile is set as the profile satisfying the temperature boundary conditions without flow, i.e.
\begin{align}
  \theta&= \begin{cases}
    h_i-z+\sum\limits_{m=1}^{4}\frac{P_m}{K_mH}[e^{-K_mHh_i}-e^{-K_mHz}], & z<h_i \\
    \frac{z-h_i}{1-h_i}\theta_c, & z\ge h_i
  \end{cases}
  \label{eq:Tprof}
\end{align}
where $z=Z/H$ is the dimensionless depth and $\theta_c=T_c/\Delta_T$ the non-dimensionalized bottom temperature. Additional details on the numerical methods can be found in the Supplementary Materials.

Given the initial condition as in eq.~(\ref{eq:Tprof}), the water can either fully freeze or partially melt. An example of the final stage is shown in fig.~\ref{fig1}. With an exponentially decaying radiation profile, the majority of heat absorption occurs near the surface. Consequently, warmer water accumulates near the top, while colder water resides at the bottom. Due to the density anomaly, the warmer water is denser than the colder water, resulting into convection. As the bottom temperature remains below 0, an equilibrium state is reached, depending on the values of $Ra\propto I_0$ and $h_i$. It is also possible for the ice to completely freeze in the final equilibrium state, i.e., the depth is zero. The achieved equilibrium state is the primary focus of this paper, and we investigate how this state is affected by $Ra\propto I_0$ and $h_i$.

Fig.~\ref{fig2}(a) presents the temporal evolution of the melt pond depth for cases with a fixed $I_0=35 W/m^2$, while varying $h_i$ from $0.1$ to $0.95$. The results demonstrate that the final equilibrium state is not unique: what state is reached depends on the initial depth $h_i$, showing the two possible outcomes of our simulations. When $h_i$ is smaller than the value of the dashed line, the ice fully refreezes, as shown in fig.~\ref{fig2}(b). Conversely, when $h_i$ exceeds this value, the melt pond either partially melts or partially freezes to reach the value of the solid line, with the flow field shown in fig.~\ref{fig2}(c). It is noteworthy that bistability also occurs for melting in classical Rayleigh-Bénard convection \cite{purseed2020bistability}, where it arises from the transition from the diffusive to the convective equilibrium, and thus is limited to relatively \textit{small} $Ra$ values. However, in contrast to that case, the bistability observed here occurs for relatively \textit{high} $Ra$ (realistic values for melt ponds), more realistic radiative heating, and typical depths of melt ponds and is thus very relevant for melt pond evolution processes.

\begin{figure}[ht]
 \centering
 \includegraphics[width=1.0\columnwidth]{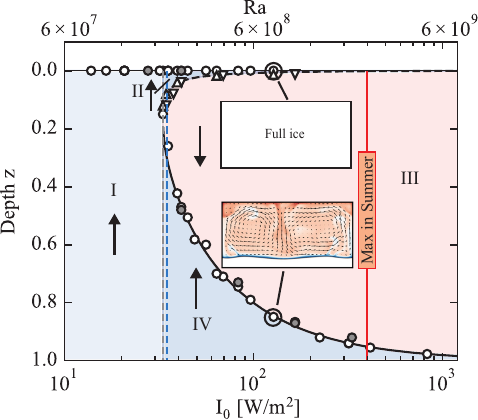}
 \caption{The bistability diagram for various $Ra$ (or $I_0$) and initial depth $h_i$ from our numerical results. The circles represent the stable final equilibrium state of the 2D (in white color) and 3D (in grey color) simulations. The pairs of triangles are starting conditions close to the bifurcation, with the left one fully freezing to $h=0$ and the right one partially melting to an equilibrium depth. The black solid (dashed) curve represents the theoretical prediction for the stable (unstable) solution from the heat flux balance (eq.~(\ref{eq:ht_balance})). The gray vertical dashed line shows the bound of regime I. The vertical red line shows the maximum solar radiation during summertime for melt ponds in the Arctic \cite{kim2018salinity}. The blue vertical dashed line shows $I_0=35W/m^2$ as in fig.~\ref{fig2}. The insets show snapshots (same color bar as in fig.~\ref{fig2}(c)) of two distinct final states with the same $Ra$ but different $h_i$. The four black vertical arrows show the direction of the ice front propagation in four different regimes: up in I, II \& IV, down in III}
 \label{fig3}
\end{figure}

Fig.~\ref{fig3} shows the phase diagram of the final equilibrium depths (circles). Zero value means full ice and non-zero values mean a melt pond with a certain depth. The phase diagram displays a bifurcation and four regimes. Regime I - for relatively small $Ra$ (left of the gray dashed line), ice always completely refreezes regardless of $h_i$. Regime II - for large enough $Ra$ but small $h_i$ (above the black dashed line), the ice also completely refreezes. Regime III - for mediate $h_i$ and large enough $Ra$, the ice partially melts down to a stable equilibrium state (the black solid line) of the melt pond. In Regime IV - for large $h_i$ and large enough $Ra$, the ice will freeze towards the stable equilibrium state. The subcritical bifurcation (or tipping) point is characterized by an abrupt transition, triggered by a small perturbation of the initial condition. The realistic value $I_0$ for the light intensity during summertime varies between $0$ and $400 W/m^2$ \cite{kim2018salinity}. This implies strong hysteresis behavior of the melting \& freezing process: as $I_0$ increases beyond a threshold, a little perturbation of $h_i$ can result in the melt pond formation, and the melt pond cannot be refrozen as $I_0$ decreases again.

In order to illustrate the mechanism of the bistability, we examine different scenarios. When the radiation intensity is insufficient, regardless of the initial water depth, the conduction of heat through the ice outweighs the radiation-driven heat flux in the water. Consequently, the ice layer continues to grow until it fills the whole domain. Conversely, in the presence of strong radiation and a thin layer of water, the heat flux in the water remains weak compared to the conduction in the ice, also leading to ice growth. Only when there is a certain amount of water, capable of absorbing sufficient radiation and generating a large heat flux, the ice will (partially) melt. As the melting progresses and the ice layer thins, the conductive heat flux increases, establishing another equilibrium point. Thus, the formation or refreezing of the melt pond depends on both the radiation intensity $I_0$ and the initial depth $h_i$.

To quantitatively describe the equilibrium states, we consider the heat flux balance at the ice front. The depth-dependent heat flux in the water $Q_w(z)$ can be calculated from the spatial integral of the time-averaged energy equation, which gives 
\begin{equation}
Q_w(z)= \Sigma P_m(1-e^{-K_mHz}).
\end{equation}
$Q_w(z)$ satisfies the adiabatic boundary condition at $z=0$, $Q_w(z=0)=0$. In the ice, the heat is transferred by pure conduction, which gives
\begin{equation}
Q_i(z)=\frac{\theta_c}{1-z},
\end{equation}
When $Q_w(z)>Q_i(z)$ at the ice surface, the ice melts. When $Q_w(z)<Q_i(z)$ at the ice surface, the ice freezes. At equilibrium depth, the heat fluxes in water and in ice should balance, $Q_w(z)=Q_i(z)$, i.e.
\begin{equation}
\Sigma P_m(1-e^{-K_mHz})=\frac{\theta_c}{1-z}.
\label{eq:ht_balance}
\end{equation}
Eq.~(\ref{eq:ht_balance}) can numerically be solved, as the intersection points of the functions $Q_i(z)$ and $Q_w(z)$. The stable (black solid curve) and unstable (black dashed curve) solutions from eq.~(\ref{eq:ht_balance}) are shown in fig.~\ref{fig3}, which coexist for large enough $Ra$. Above the unstable solution, the ice freezes, while descending and crossing the unstable solution leads to abrupt partial ice melting until the equilibrium depth from the stable solution is reached. In the case of low $Ra$, no solution exists, indicating complete freezing of the ice, irrespective of the initial conditions. The equilibrium depth solution obtained from the heat flux balance exhibits excellent agreement with our numerical results. For different bottom ice temperatures $\theta_c\ne -3^oC$ \cite{bliss2020glaciers}, based on our theoretical analysis (eq.~(\ref{eq:ht_balance})), we anticipate the persistence of bistability, albeit with a horizontal shift in the phase diagram.

\begin{figure}[ht]
 \centering
 \includegraphics[width=1.0\columnwidth]{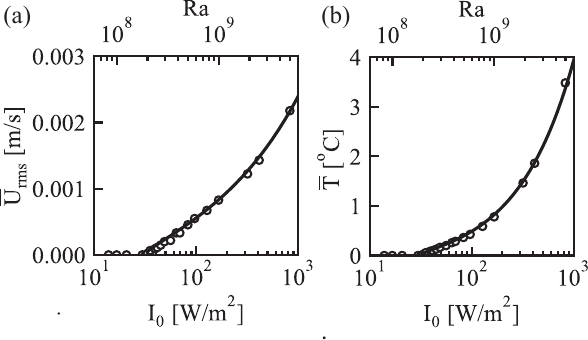}
 \caption{(a) The bulk-averaged root-mean-square velocity $\overline{U}_{rms}$ as a function of $I_0$. (b) The bulk-averaged temperature $\overline{T}$ as a function of $I_0$. The solid lines represent the prediction curve from GL theory (eq.~(\ref{eq:GL2a})-(\ref{eq:GL2b})), with the prefactors obtained by a least-squares fitting.}
 \label{fig4}
\end{figure}

How do the flow strength and the bulk temperature inside melt ponds change with increasing radiation intensity? We quantify this in terms of a bulk-averaged root-mean-square (rms) velocity, $\overline{U}_{rms}$ (dimensional), and the bulk-averaged temperature $\overline{T}$ (dimensional). Both are given in fig.~\ref{fig4} as a function of $Ra$. At low $Ra$, both these quantities are zero since the system solely consists of ice. However, as $Ra$ increases, both $\overline{U}_{rms}$ and $\overline{T}$ experience a noticeable rise.

To make predictions for the $Ra$-dependence (and thus radiation intensity $I_0$-dependence) of these quantities, we apply the Grossmann-Lohse (GL) theory \cite{grossmann2000scaling,grossmann2001thermal}, in its extension to uniform internal heating \cite{wang2021scaling}. This theory is based on the splitting of the viscous and thermal dissipation rates in boundary and bulk contributions, respectively. We apply the scaling relations from this extended theory for radiative heating using the exponentially decaying profile described in eq.~(\ref{eq:profile}) (Further details of the derivation are provided in the Supplementary Materials). For the bulk-averaged temperature $\overline{T}$ and the bulk-averaged Reynolds number $Re=H\overline{U}_{rms}/\nu$, they read \cite{wang2021scaling}:
\begin{equation}
\overline{T}/\Delta_T\sim Ra^{-\frac{1}{5}},\ Re\sim Ra^{\frac{1}{2}},\ \overline{U}_{rms}/U_f\sim Ra^0,
\label{eq:GL1}
\end{equation}
where $U_{f}=\sqrt{\alpha gI_0H^2/(\rho c\kappa)}$ is the characteristic velocity. In our analysis, we consider the effective $Ra$ and $Re$ with the length scales defined as the dimensional equilibrium depth $H_0$ of the melt pond, obtained from eq.~(\ref{eq:ht_balance}), which is a function of $I_0$. This enables us to establish the dependence of $\overline{T}$ and $\overline{U}_{rms}$ on $Ra$ based on eq.~(\ref{eq:GL1}) as:
\begin{align}
& \overline{T} \sim Ra_0^{-\frac{1}{5}}\Delta_0\sim \left(Ra\left(\frac{H_0}{H}\right)^4\right)^{-\frac{1}{5}}\left(\left(\frac{H_0}{H}\right)^2\Delta_T\right)\sim I_0^{\frac{4}{5}}H_0^{\frac{6}{5}}, \label{eq:GL2a}\\
& \overline{U}_{rms} \sim U_{f_0}\sim I_0^{\frac{1}{2}}H_0^{\frac{3}{2}}, \label{eq:GL2b}
\end{align}
where the characteristic velocity $U_{f_0}$ and characteristic temperature $\Delta_0$ are determined by the equilibrium depth $H_0$. We use subscript $0$ to represent these quantities defined by $H_0$. The final predictive curves are shown as black solid lines in fig.~\ref{fig4}, and they all align very well with the numerical data. For the highest $I_0$, $\overline{T}$ is about 4 degrees, close to the temperature of maximal density, so that there will be some non-Oberbeck–Boussinesq effects here. These are not captured by the model, which however works for the current situation in polar regions \cite{kim2018salinity}, where the temperatures are smaller and the density approximately linearly depends on temperature.

In conclusion, our numerical investigation sheds light on the intricate process of melt pond formation under the influence of solar radiative heating. We have revealed the occurrence of bistability, wherein the melt pond formation depends on the initial conditions and the solar radiation strength. This bistability represents a tipping point in the realm of climate science, as even a small perturbation can lead to an abrupt transition. Based on the heat flux balance between water and ice, a physical and quantitative understanding of the bistability behavior is given, which shows good agreement with the numerical results. Using the GL theory, we can further quantitatively account for the dependence of the bulk-averaged temperature and flow velocity on the solar radiation strength.

Note that the model we used is highly idealized, to make the model more realistic in the next step, surface heat flux exchange with the cold air could be incorporated, along with depth-dependent albedo and the effect of surface waves on shallow melt pond, the transmission of solar radiation into the ice, and the reflection/absorption from the bottom surface \cite{skyllingstad2007numerical}. Considering the latter will modify the heat flux profile in the heat flux balance equation (eq.~\ref{eq:ht_balance}), though the bistability will still exist; the stable and unstable equilibria will only slightly be shifted. We have estimated that the contribution from the reflection \cite{skyllingstad2007numerical} is $\sim0.1$ times that of the solar influx for a melt pond with a depth of $0.2m$. Furthermore, it is worth exploring the effect of salinity on the melt pond dynamics \cite{kim2018salinity,yang2023salinity}. Then complex mushy layer structures can form \cite{worster1997convection} and brine convection can occur inside \cite{du2023sea}. By considering these factors and refining our models, we can better understand and model the melt pond evolution process.

\textit{Acknowledgement} We acknowledge PRACE for awarding us access to MareNostrum in Spain at the Barcelona Computing Center (BSC) under the project 2020235589 and project 2021250115. We also acknowledge support by the German Science Foundation DFG through the Priority Programme SPP 1881 ''Turbulent Superstructures'' and by the ERC Advanced Grant under project “MultiMelt” with No. 101094492.

\end{document}